\RequirePackage{rotating}
\documentclass[twocolumn,useAMS,usenatbib, 12pt]{aastex6}
\usepackage{graphicx, subfigure}

\usepackage{lipsum}
\usepackage{array,booktabs,dcolumn,calc}
\usepackage[fleqn]{amsmath}
\usepackage{float}
\usepackage{bm}
\usepackage{amssymb,amsmath}
\usepackage{setspace}
\usepackage{rotating}
\usepackage{xcolor}
\hypersetup{colorlinks = true, citecolor=black, linkcolor=blue}

\begin{document}

\title{Measuring filament orientation: a new quantitative, local approach}

\author{C.-E. Green\altaffilmark{1,2}, J.~R. Dawson\altaffilmark{2,3}, M.~R. Cunningham\altaffilmark{1}, P.~A. Jones\altaffilmark{1}, G. Novak\altaffilmark{4}, L.~M. Fissel\altaffilmark{5}}

\altaffiltext{1}{School of Physics, University of New South Wales, Sydney, NSW, 2052, Australia}
\altaffiltext{2}{CSIRO Astronomy \& Space Science, Australia Telescope National Facility, PO Box 76, Epping, NSW 1710, Australia}
\altaffiltext{3}{Department of Physics and Astronomy and MQ Research Centre in Astronomy, Astrophysics and Astrophotonics, Macquarie University, NSW 2109, Australia}
\altaffiltext{4}{Center for Interdisciplinary Exploration and Research in Astrophysics (CIERA) and Department\ of Physics \& Astronomy, Northwestern University, 2145 Sheridan Road, Evanston, IL 60208, U.S.A.}
\altaffiltext{5}{National Radio Astronomy Observatory (NRAO), 520 Edgemont Road, Charlottesville, VA, 22903}

\begin{abstract}
The relative orientation between filamentary structures in molecular clouds and the ambient magnetic field provides insight into filament formation and stability. To calculate the relative orientation, a measurement of filament orientation is first required. We propose a new method to calculate the orientation of the one pixel wide filament skeleton that is output by filament identification algorithms such as \textsc{filfinder}. We derive the local filament orientation from the direction of the intensity gradient in the skeleton image using the Sobel filter and a few simple post-processing steps. We call this the `Sobel-gradient method'. The resulting filament orientation map can be compared quantitatively on a local scale with the magnetic field orientation map to then find the relative orientation of the filament with respect to the magnetic field at each point along the filament. It can also be used in constructing radial profiles for filament width fitting. The proposed method facilitates automation in analysis of filament skeletons, which is imperative in this era of `big data'. 

\end{abstract}

\keywords{
ISM: structure --- 
methods: statistical ---
methods: data analysis ---
stars: formation ---
techniques: image processing}
\section{Introduction} \label{sec:intro}
Elongated filaments of gas and dust are ubiquitous in molecular clouds \citep[e.g.][]{molinari2010}. These clouds are stellar nurseries and the filaments they host may play an important role in star formation, with the majority of star-forming cores lying along filaments ``like beads on a string" \citep{andre2014}. \\

Filaments represent velocity coherent over-densities of gas and dust, and have aspect ratios greater than at least three \citep{panopoulou2014}. They can be identified from a 2D astronomical image, for instance a column density map, using skeleton-based filament identification algorithms such as \textsc{filfinder} \citep{koch2015}. For a given input image and set of input parameters these return a filament skeleton. The skeleton is a one pixel wide representation of the filamentary structure in the original image, tracing the main path of the filament and its branches. Clumps and cores are also over-dense compared to their surroundings, and are distinguished from filaments by smaller aspect ratios of $\sim$2 \citep{tachihara2000}. Clumps are inhomogeneously dense velocity coherent regions from which a system of stars may form. A core is a dense velocity coherent region that may form a single star or binary star. Cores are usually found grouped into clumps.\\

In the study of these filaments, one useful measurement is that of their orientation. Filament orientation is used in the construction of radial profiles used to derive filament width. Filament orientation can also be compared with that of the magnetic field. Magnetic fields are believed to have a dynamically important role in filament formation and stability. In several theories of cloud structure formation matter is channelled along the field lines, allowing filaments to form through gravitational contraction \citep{nakamura2008}. In this scenario dense filaments would be aligned perpendicular to the field and less dense filaments would be aligned parallel \citep{li2008}. \citet{goldsmith2008} and \citet{planck2016} find observational evidence for this scenario. \\

\begin{figure}[t]
       \centering
       \caption{\textbf{Gradient vector for an ideal edge.} A grey line bounds the image, but is not part of this example. The gradient vector is perpendicular to the ideal edge (transition from black to white) and points towards the higher intensity (`lighter') values (where black=0, and white=255). \label{fig:grad_eg}}

       \includegraphics[width=0.25\textwidth, clip=true, trim=0cm 0cm 0cm 0cm]{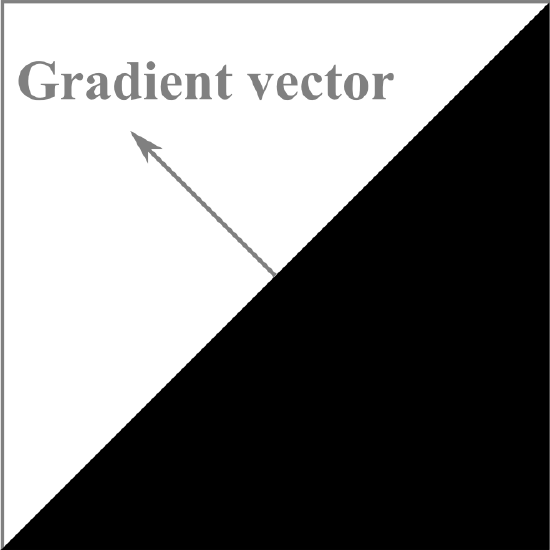}
\end{figure}

Filament orientation can be measured from the filament skeleton \citep{koch2015}. The intensity changes at the edge of the skeleton, and this intensity gradient has an associated direction (see~\autoref{fig:grad_eg}). Here we propose a new method to derive the filament orientation, exploiting this fact. This is achieved through the use of the Sobel filter, described in~\autoref{sec:fil_orient}, and some additional post-processing steps discussed in~\autoref{sec:post}. This method, which we call the `Sobel-gradient method', returns a quantitative and local map of filament orientation for any filament skeleton, including those with complex interconnected structures\footnote{The associated \textsc{python} code and documentation will be available on \textsc{github} in the near future. In the meantime please contact the author for an early release.}. The map reveals how the orientation changes as the skeleton curves on a local scale. We explore the uncertainties associated with the method in~\autoref{sec:test_suite}. Applications for this method are suggested in~\autoref{sec:applications}. \\

\subsection{Motivation}
There are two main existing quantitative approaches to measuring filament orientation, the first being a map based analysis (e.g. \citealt{schisano2014} Hessian matrix method), and the second being a skeleton based analysis (e.g. \citealt{koch2015} \textsc{filfinder} algorithm). Prior to their introduction the predominantly utilized method for measuring filament and field relative orientation, was a qualitative, global, visual comparison \citep[e.g.][]{goldsmith2008, busquet2013, palmeirim2013}, and this approach is still used in more recent works \citep[e.g.][]{kusune2016}.\\ 

The \citet{schisano2014} method uses the Hessian matrix to identify filaments and measure their orientation from a 2D astronomical map such as a Herschel\footnote{Herschel is an ESA space observatory with science instruments provided by European-led Principal Investigator consortia and with important participation from NASA.} \citep{pilbratt2010} dust column density map. However \citet{schisano2014} state that ``for very complex features where filaments are organized in web-like structures, the cross-spine profile fitting often fails to converge". In our related project studying the filamentary structure of the South- (SR) and Centre-ridges (CR) of the Vela C Molecular Cloud (C.-E. Green et al. 2017, in preparation), we tried this method to derive the orientation of the filaments shown in~\autoref{fig:sobel} panel (a). Indeed, the method failed to converge for this complex interconnecting data-set, motivating our search for an alternate method. \\

The \citet{koch2015} filament identification algorithm \textsc{filfinder} has an inbuilt skeleton-based filament orientation calculator that uses a `line-based' approach, the Rolling-Hough Transform (RHT). In this type of approach test lines with different angles are fit to groups of pixels along the filament skeleton, finding the best fit line and thus the associated angle of the skeleton segments. \citet{koch2015} define filament orientation to be the weighted directional mean of the distribution of angles for the skeleton returned by the RHT. \textsc{filfinder} thus returns to the user a single orientation value over long filament segments (e.g. $\sim$40 pixels, which corresponds to 1.2\,pc in our example Vela C SR data in~\autoref{fig:sobel} panel (a)), whereas our more complex filaments curve on a smaller scale of $\sim$5\,pixels (0.15\,pc). This definition of filament orientation is therefore not compatible with our goal of a quantitative, local, ``position-by-position" filament orientation.\\

This motivated our search for an alternate, fully automated filament orientation measurement method that:
\begin{enumerate}
\item returns a quantitative, local, ``position-by-position" measurement of filament orientation,

\item can be applied to complex interconnecting filaments (such as the SR shown in~\autoref{fig:sobel} panel (a)), as well as simpler, more linear filaments.
\end{enumerate} 

As previous map based approaches such as \citet{schisano2014}'s Hessian matrix method have not worked for these more complicated `looped' (in the 2D image) filaments we focussed our search on a skeleton based approach that would provide a measurement on a smaller scale than the RHT method built into \textsc{filfinder}. This led us to develop the new Sobel-gradient method we propose here, exploiting the image intensity gradient to arrive at a map of filament orientation.  \\

\section{Filament orientation from the image intensity gradient} \label{sec:fil_orient}
Filament skeletons are generally output by filament identification algorithms in Flexible Image Transport System (FITS\footnote{\href{http://fits.gsfc.nasa.gov/fits\textunderscore primer.html}{http://fits.gsfc.nasa.gov/fits\textunderscore primer.html}}) file format where the pixels `on' the skeleton have a value of one, and those `off' the skeleton have a value of zero. These can be trivially converted to a greyscale image matrix, where the `on' skeleton pixels are white, with a value of 255 and the `off' skeleton pixels are black, with a value of zero.  We use the \textsc{python scipy ndimage} implementation of the Sobel filter where it is necessary to use this convention. White could also be represented as a value of one, black as a value of zero and grey shades as decimal values in between, if a different implementation was used. As the skeleton is a binary image, the image intensity gradient only exists at the edges of the filament skeleton where the intensity changes. This intensity gradient has a magnitude and a direction. An example of the image intensity gradient vector for an idealised case is shown in~\autoref{fig:grad_eg}. With some minor adjustments the skeleton orientation can be derived from the intensity gradient direction. \\

\subsection{Intensity gradient direction} 
\label{sec:grad_direc}
To calculate the direction of the intensity gradient, we need the first $x$ and $y$ derivatives ($G_{x}$ and $G_{y}$ respectively) of the skeleton image matrix, $I$.  The direction, $\Theta$, of the gradient is calculated as:
\begin{equation}
\label{eqn:grad_direc}
\Theta=tan^{-1}(G_{y}/G_{x})
\end{equation}

The Sobel filter is commonly used in computer vision to estimate these derivatives \citep{gao2010}. It is already built into \textsc{matlab} and the \textsc{python scipy ndimage} library so this method can be quickly and easily implemented. The Sobel filter itself is computationally inexpensive. Its speed is a major advantage because orientation measurements are generally repeated for the multiple different skeletons produced by different combinations of input parameters to the filament identification algorithm. In some ways this approach is similar to the Histogram of Relative Orientations (HRO) method of \citet{soler2013}, which also uses Gaussian derivatives (of which the Sobel filter is one of the simplest types) to measure the orientation of molecular cloud structure. Our approach differs in that we aim to find the orientation only of strictly defined and identified filaments, by measuring the orientation of the one pixel wide filament skeleton. This is in contrast to the HRO method, which makes no structure definitions, and involves finding the orientation of all structures of all scales within a column density map.\\

\subsubsection{The Sobel filter} 
\label{sec:sobel}
The Sobel filter is a discrete differential operator consisting of two 3$\times$3 matrices of coefficients. When convolved with the image matrix, $I$, two new image matrices are created, representing estimates of $G_{x}$ and $G_{y}$ as follows (where $\ast$ represents the convolution operation) \citep{gao2010}:
\begin{equation}
G_{x} = 
\left[\begin{array}{ccc} -1 & 0 & +1\\ -2 & 0 & +2\\ -1 & 0 & +1 \\ \end{array}\right]
\ast I
\label{eq:gx}
\end{equation}

\begin{equation}
G_{y} = 
\left[\begin{array}{ccc} +1 & +2 & +1\\ 0 & 0 & 0\\ -1 & -2 & -1 \\ \end{array}\right]
\ast I
\label{eq:gy}
\end{equation}

Angular measurements have a reference point and a direction of increase. For the image gradient returned by the Sobel filter, these are the horizontal and the anticlockwise direction. Therefore in that convention the gradient angle needs to be rotated by 90$^{\circ}$ to give the angle of the edge. However in the astronomical convention the reference point is North (often vertically upwards in astronomical images), with an anticlockwise direction of increase. We are operating in the domain of [90, -90] so we therefore only need to perform a simple sign reversal  to arrive at an estimate of the orientation of the skeleton edge\footnote{North is vertically upwards for the Vela C data presented in this work. If this is not the case for the users dataset this step would require the relevant rotation and sign adjustment to account for that.} when working in the astronomical convention. \\ 

\section{Deriving the skeleton orientation}
\label{sec:post}
\begin{figure*}[t]
       \centering
       \caption{\textbf{Deriving intensity gradient direction for a filament skeleton with the Sobel filter.} Panel (a) shows the image input to \textsc{filfinder}, a Herschel dust column density map of the Vela C South-ridge of \citet{fissel2016}. Panel (b) is the skeleton output by \textsc{filfinder} for the South-ridge, and panel (c) is that skeleton with its junctions removed. These three panels have had their colors inverted for easier viewing. For the purposes of illustration, panels (d), (e) and (f) show quantities that were calculated separately for each connected component of the skeleton (see discussion in text), but have been plotted together. The Sobel filter is applied to the each connected component of the skeleton, producing the $x$ and $y$ derivatives ($G_{x}$ and $G_{y}$) shown in panels (d) and (e) respectively. The direction of the image intensity gradient is then calcuated for each component using~\autoref{eqn:grad_direc}, and is plotted in panel (f). The sign of that map is reversed for consistency with astronomical conventions. The grey does not form part of the colourmaps in these panels. It shows the `Not-a-Number' (NAN) background, as white forms part of the colourmap used. The $x$ and $y$ axes are plotted in pixel coordinates where the lower left is the origin. These images have been zoomed to the region $x$=55-226, $y$=65-265, for easier viewing. \label{fig:sobel}}
       \includegraphics[width=0.83\textwidth, clip=true, trim=1cm 0.5cm 1cm 0.5cm]{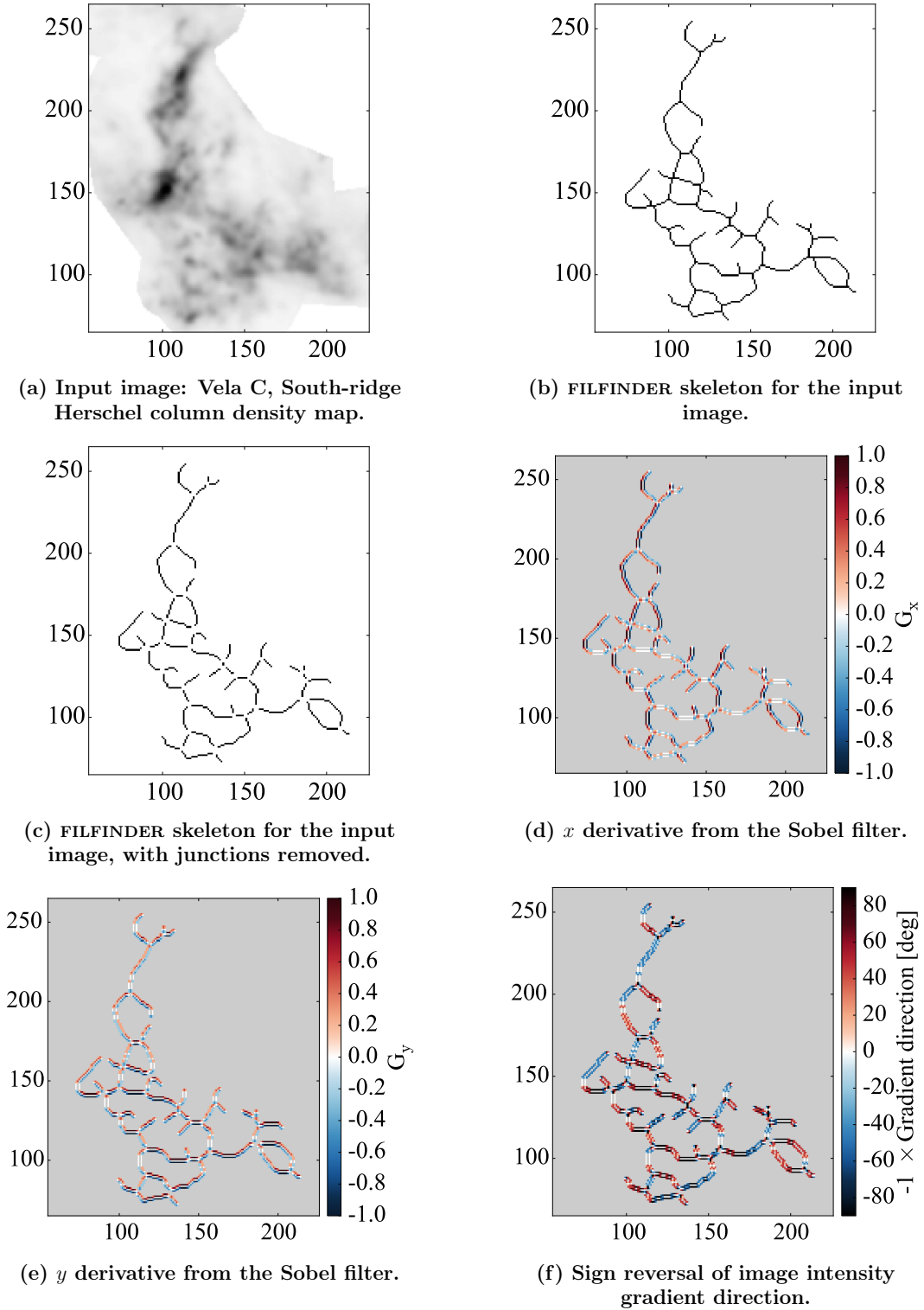}
       
\end{figure*}

Throughout this work we will use, for the purpose of illustration, filament skeletons identified with \textsc{filfinder} from \citet{fissel2016} Herschel dust column density images of the SR and CR of Vela C. For the SR these are shown in~\autoref{fig:sobel} panels (a) and (b) respectively. In the SR and CR data one pixel corresponds to 0.03\,pc. The skeletons were selected as belonging to the group of optimum skeletons, most similar to, and therefore the best representation of, the original column density image, using the mean structural similarity index as a goodness-of-fit measure as described in \citet{green2017}. Together the selected skeletons from the SR and CR comprise a representative data set, containing interconnected `loops' and curvature on the small scale along with some more linear segments. They were selected as they contained the largest number of `difficult' features for the algorithm to tackle. The SR skeleton selected was produced by \textsc{filfinder} input parameters of: skeleton threshold (skeleton length cutoff) of 10 pixels (0.3\,pc, corresponding to an aspect ratio of 3 \citep{panopoulou2014}, given an assumed width of 0.1\,pc \citep{arzoumanian2011}), branch threshold (branch length cutoff) of 3 pixels (0.09\,pc), global threshold (noise threshold) of 69\%, flattening threshold (threshold for arctan flattening which removes impact of compact sources like clumps in masking step) of 60\%. The CR skeleton was produced by a skeleton threshold of 10 pixels (0.3\,pc), a branch threshold of 5 pixels (0.15\,pc), a global threshold of 74\%, and a flattening threshold of 96\%. \\

Before deriving the filament skeleton orientation we first automatically remove junction points\footnote{Junction points are locations where filaments meet, i.e. locations where an on-skeleton pixel has more than two on-skeleton pixel neighbours. We define neighbours as the eight pixel positions surrounding a central pixel enclosed within a 3$\times$3 window.} from the skeleton. Junction points belong to all of the intersecting filaments involved and therefore have an undefined orientation. Removing these breaks up the skeleton into many components as shown in~\autoref{fig:sobel} panel (c), which in computer vision are called connected components. The image gradient is defined at the pixels immediately surrounding the skeleton, therefore the gradients may overlap and overwrite each other at the new endpoints created by deleting the junctions since they are so close together. To avoid this issue we automatically locate and label the connected components\footnote{Connected components labelling algorithms exist in many computing languages. They can be labelled automatically using e.g. \textsc{python's} \textsc{ConnectedComponentsWithStats} function from the \textsc{OpenCV} library.} and repeat the Sobel-gradient method described in the following on each component separately, collecting the final orientation maps of each component into a `master map', the final filament skeleton orientation map.\\

To derive the skeleton orientation of each component we calculate the $x$ and $y$ image derivatives using the Sobel filter (Equations \ref{eq:gx} and \ref{eq:gy} repectively), and then calculate the skeleton image gradient using~\autoref{eqn:grad_direc}. The sign of the skeleton image gradient is then reversed for consistency with astronomical conventions. This process is illustrated in~\autoref{fig:sobel} panels (d), (e), and (f). \\

\begin{figure*}[t]
       \centering
       \caption{\textbf{Post-processing to move from intensity gradient direction to skeleton orientation.} For the purposes of illustration, all panels show quantities that were calculated separately for each connected component of the skeleton (see discussion in text), but have been plotted together. Panel (a) illustrates the two issues that need to be resolved to move from a map of intensity gradient direction to that of skeleton orientation. This image is an annotated version of that shown in~\autoref{fig:sobel} panel (f). The pixels at filament ends are deleted and replaced with the circular vector average of the values of the nearest unaffected pixels in the filament, giving panel (b). Then the blank central pixels are infilled with the circular vector average of their neighbours in panel (c). Circular vector averaging is performed smoothing the map, resulting in panel (d), a map of the filament skeleton orientation.  Panel (d) is the `master map' to which the skeleton orientation for each connected component is saved. The grey shows the `Not-a-Number' (NAN) background. The $x$ and $y$ axes are plotted in pixel coordinates where the lower left is the origin. These images have been zoomed to the region $x$=55-226, $y$=65-265, for easier viewing. \label{fig:post}}
        \includegraphics[width=0.99\textwidth, clip=true, trim=0.5cm 4cm 0.5cm 3.5cm]{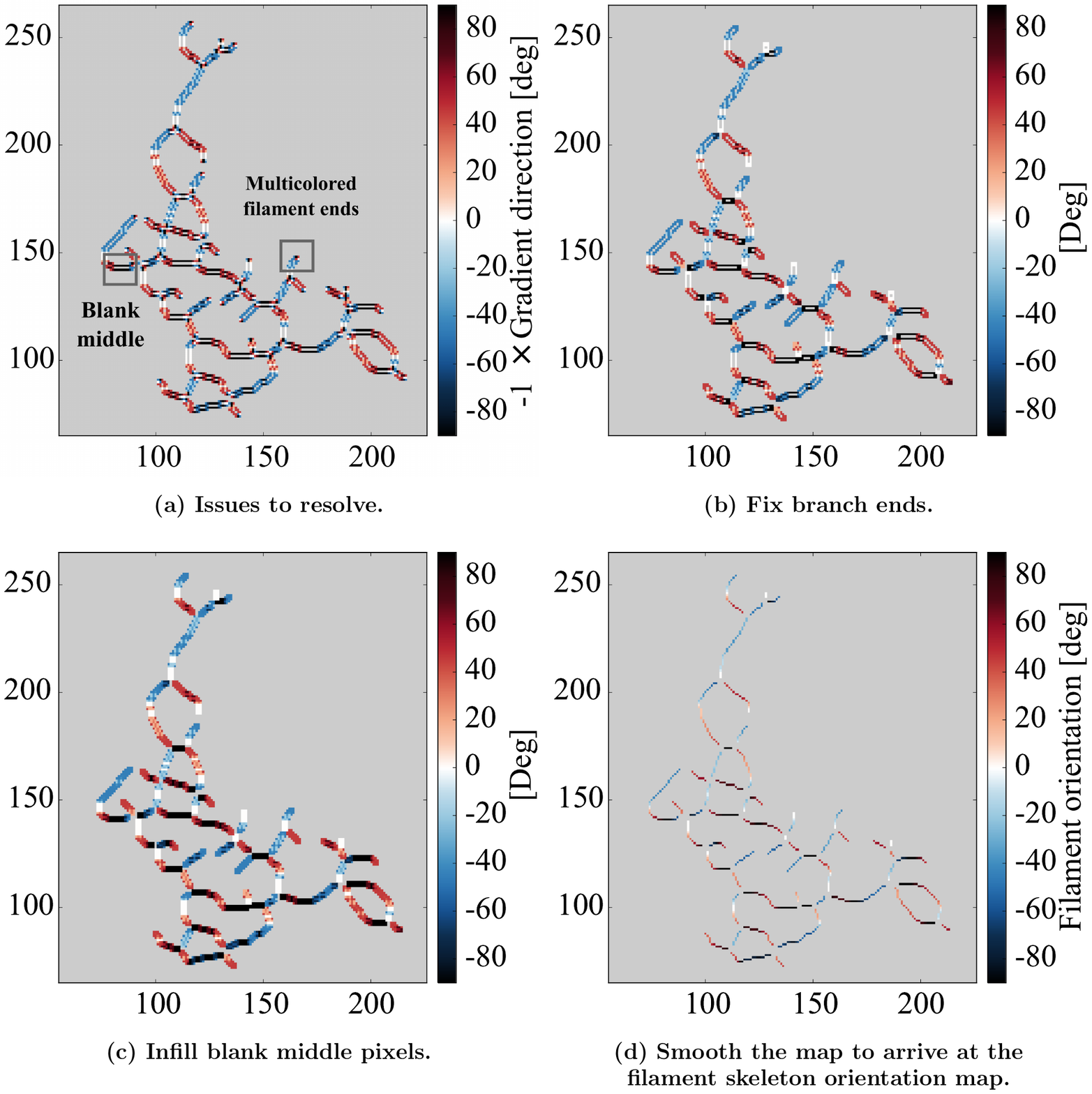}

\end{figure*}

The sign reversed gradient direction map in~\autoref{fig:sobel} panel (f) gives the orientation of the edge of the skeleton. To move from this to a map of the orientation of the skeleton's path we perform some simple post processing steps, which are demonstrated in~\autoref{fig:post}. These correct minor, partially cosmetic issues that arise as a direct consequence of the nature of the Sobel filter image gradient approach. In these steps we 1) correct the branch ends, 2) infill the centre pixels, 3) smooth the map, and 4) select the orientation values at the positions along the original skeleton to save into the master map. These steps are illustrated in~\autoref{fig:post}. \\

Firstly, at the ends of the connected components there are pixels with angle values that are roughly orthogonal to the rest of the component. This is because of the additional exposed pixel edges at the component ends. We automatically detect and delete\footnote{There are a set number of patterns of off- and on-skeleton pixels that can occur around a component end that we test against to detect them.} the handful of affected pixels at component ends in the map. We then give them the value of the circular vector average of the closest\footnote{The closest pixels are the pixels within a 3$\times$3 window centred on the `bad' pixel closest to the rest of the filament, whose value was removed in the previous step.} unaffected values in the branch, resolving the issue as shown in~\autoref{fig:post} panel (b). Wherever angles undergo averaging we use circular vector averaging. This ensures angles are averaged correctly, accounting for the fact that they are a circular quantity that wraps back around such that e.g. 90$^{\circ}$ and -90$^{\circ}$ are equal and represent the horizontal in the astronomical definition if North is vertical.\\

Secondly, the intensity gradient only exists at the edge of the skeleton, thus leaving a partially blank centre\footnote{The centre is not entirely blank. Some of these centre pixels are `colored in' with a gradient direction value, but that value corresponds to the pixel `next door'. This occurs in skeleton sections that are not horizontal, vertical or diagonal due to the nature of the convolution of the Sobel filter with the original skeleton image.}. We automatically infill the pixels along the positions of the original skeleton with the circular vector average of their neighbouring pixels within a 5$\times$5 pixel window (also including the centre pixel in the average if it is not blank). The result is shown in ~\autoref{fig:post} panel (c). \\

Finally, we smooth the map. A 5$\times$5 pixel window is passed over the image and we consider only pixel positions lying on the original skeleton which we centre in the window. The circular vector average of the pixels inside the window is calculated and is saved to the corresponding pixel position of the central pixel in a new map. This results in a smoothed orientation map for that connected component. The smoothed map for each connected component is saved into the master map. After repeating the process for each component we arrive at a map of the filament skeleton's orientation as illustrated in~\autoref{fig:post} panel (d). \\

\begin{figure*}[t]
       \centering
       \caption{\textbf{Orientation maps.} The Sobel-gradient orientation maps for the Vela C South- and Centre-ridges. The grey shows the `Not-a-Number' (NAN) background. The $x$ and $y$ axes are plotted in pixel coordinates where the lower left is the origin. The South-ridge image in panel (a) has been zoomed to the region $x$=55-226, $y$=65-265, for easier viewing. \label{fig:example_orient}}
        \includegraphics[width=0.7\textwidth, clip=true, trim=7.3cm 8cm 7.3cm 7.5cm]{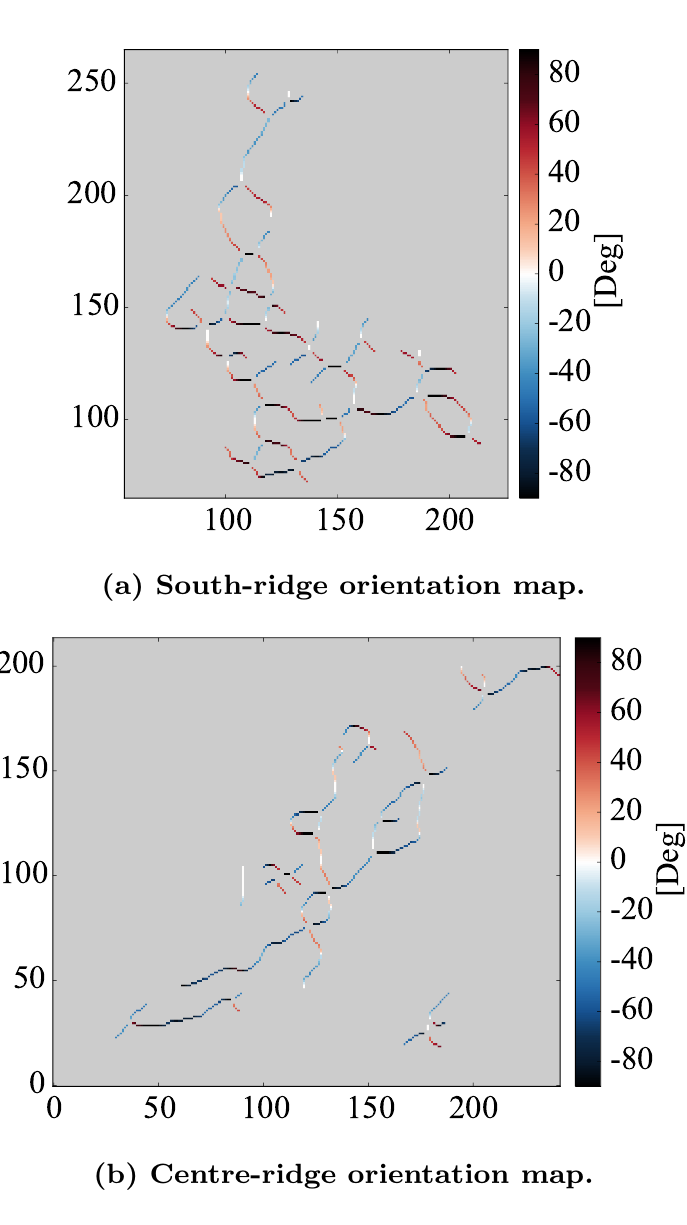}
        
\end{figure*}

These three post processing steps take us from the map of the gradient direction to a quantitative skeleton orientation map that reveals how the orientation changes as the skeleton curves on a local scale. For the first time we present the filament orientation for the SR and CR of Vela C on this small scale in  ~\autoref{fig:example_orient}. \\

\section{Constraining uncertainties}
\label{sec:test_suite}

We validate the Sobel-gradient method against the known analytic case of the circle. We generate circles with radii of 7 to 500 pixels and calculate the theoretical orientation of their tangents at each point around them. We then apply the Sobel-gradient method to them and compare the orientations at each point around the circle. Circles drawn digitally with radii smaller than 7 pixels are essentially squares with the middle pixel along each side pushed out by one position. We therefore only consider radii larger than this. \\

\begin{figure}[t]
       \centering
       \caption{\textbf{Circle test orientation differences.} The maximum, mean, and standard deviation (STD) of the differences between the Sobel-gradient and theoretical tangent orientation maps for circles of different radii. \label{fig:circle_test}}
        \includegraphics[width=0.45\textwidth, clip=true, trim=0cm 0cm 0cm 0cm]{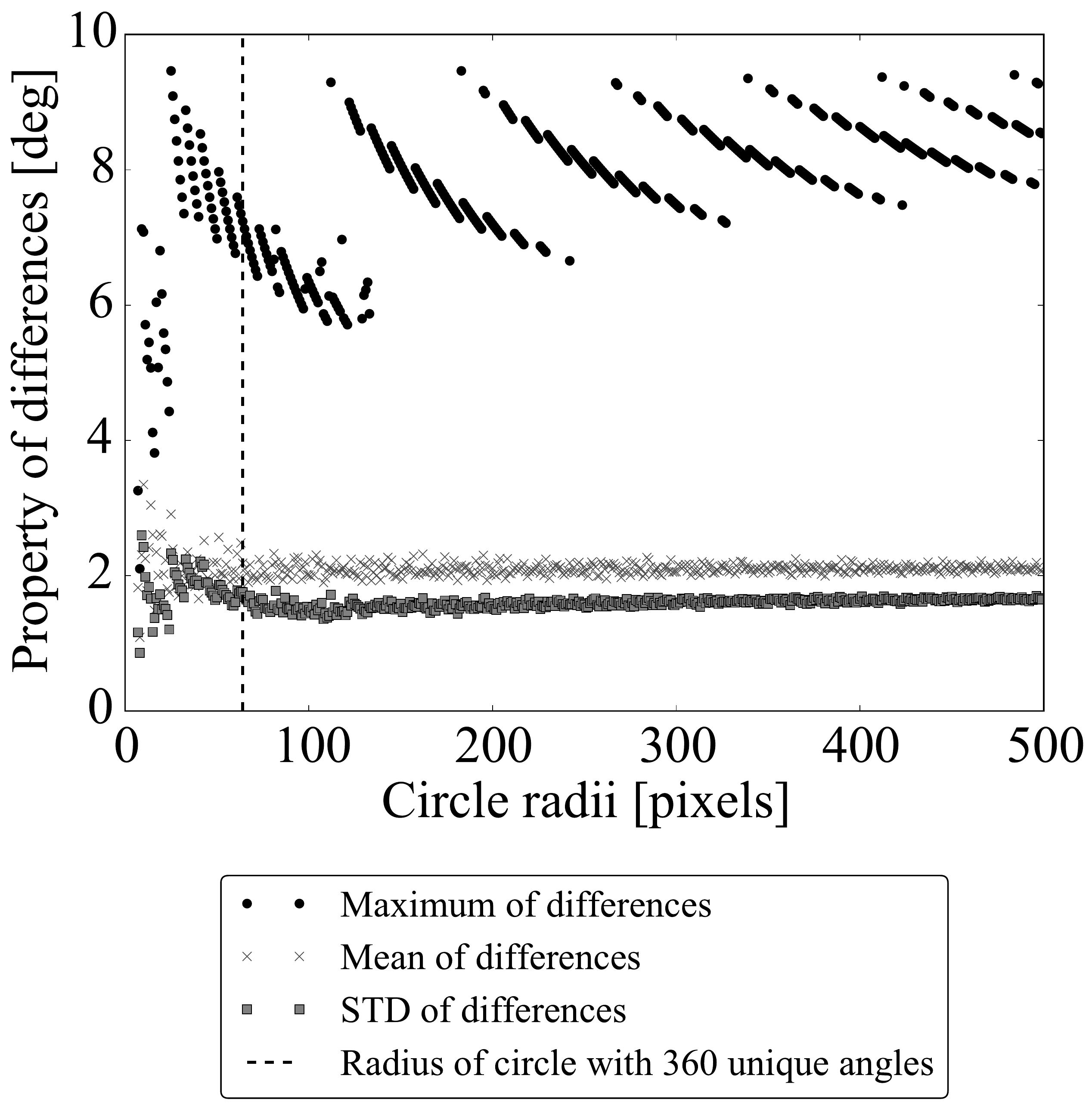}
     \end{figure}

For each radius we calculate the difference between the theoretical tangent orientation and Sobel-gradient orientation at each point around the circle, and then find the maximum, mean and standard deviation of those differences. These are plotted against their corresponding radii in~\autoref{fig:circle_test}. The average of the maximum differences at each radius was 7.8$^{\circ}$, the average of the  mean was 2.1$^{\circ}$, and the average of the standard deviations was 1.6$^{\circ}$. In calculating these values we include only the circles with radii of 57 pixels or greater, as these have 360 pixels (and therefore 360 unique angles) around them. Circles digitally generated with radii less than this are still very effected by the `squaring effect' present at small radii. This means their theoretical orientation deviates greatly from the orientation that is actually drawn. They are thus not a reliable or accurate point of validation.\\

The maximum differences are dominated by digitisation errors, which we estimate to be up to $\sim$5$^{\circ}$. The average of the standard deviations of the differences between the theoretical and Sobel-gradient orientation for each radii thus provides a more appropriate estimate of the uncertainty associated with the Sobel-gradient method. Consequently we estimate the uncertainty of the Sobel-gradient method to be $\sim$2$^{\circ}$ based on this circle test analysis.\\

Obviously circles have no start or endpoints, have no branches, and do not wiggle back and forth, changing their direction as real filaments do. To further gauge the uncertainty associated with the Sobel-gradient method in a realistic scenario we compare the Sobel-gradient orientation maps of two filamentary regions, the Vela C SR and CR, to those measured manually\footnote{Manual orientation measurements were made with a protractor on enlarged skeletons printed on paper. The estimated uncertainty of the manual measurement method is $\sim$5$^{\circ}$. This is taken as a maximum estimate, most individual manual measurements had uncertainties much smaller than this. This value includes the uncertainty of the protractor measurement, the uncertainty of decomposing the skeleton into sections, and the $\pm$1\,pixel uncertainty in the skeleton (E. Koch, 2017, private communication).}.  \\

\begin{figure*}[t]
       \centering
        \caption{\textbf{Manual orientation difference maps.} The difference between the Sobel-gradient and manual orientation map for the Vela C South- and Centre-ridges. The $x$ and $y$ axes are plotted in pixel coordinates where the lower left is the origin. The South-ridge image in panel (a) has been zoomed to the region $x$=55-226, $y$=65-265, for easier viewing. \label{fig:difference_map}}
        \includegraphics[width=0.7\textwidth, clip=true, trim=7.3cm 8cm 7.3cm 7.5cm]{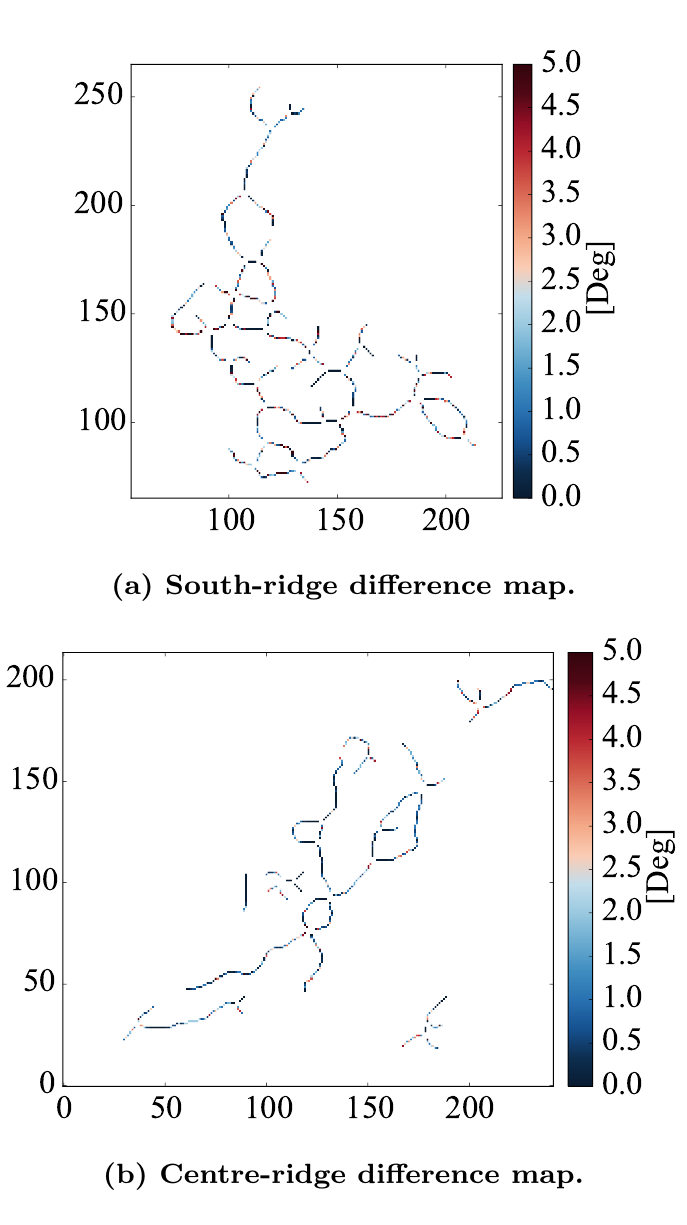}
 \end{figure*}

\begin{figure}[t]
       \centering
       \caption{\textbf{Manual orientation difference histogram.} The histograms of the difference between the Sobel-gradient and manual orientation maps for the Vela C South- (SR) and Centre-ridges (CR). Each bin is labelled with its corresponding count for each region.  \label{fig:difference_hist}}
        \includegraphics[width=0.45\textwidth, clip=true, trim=0cm 0cm 0cm 0cm]{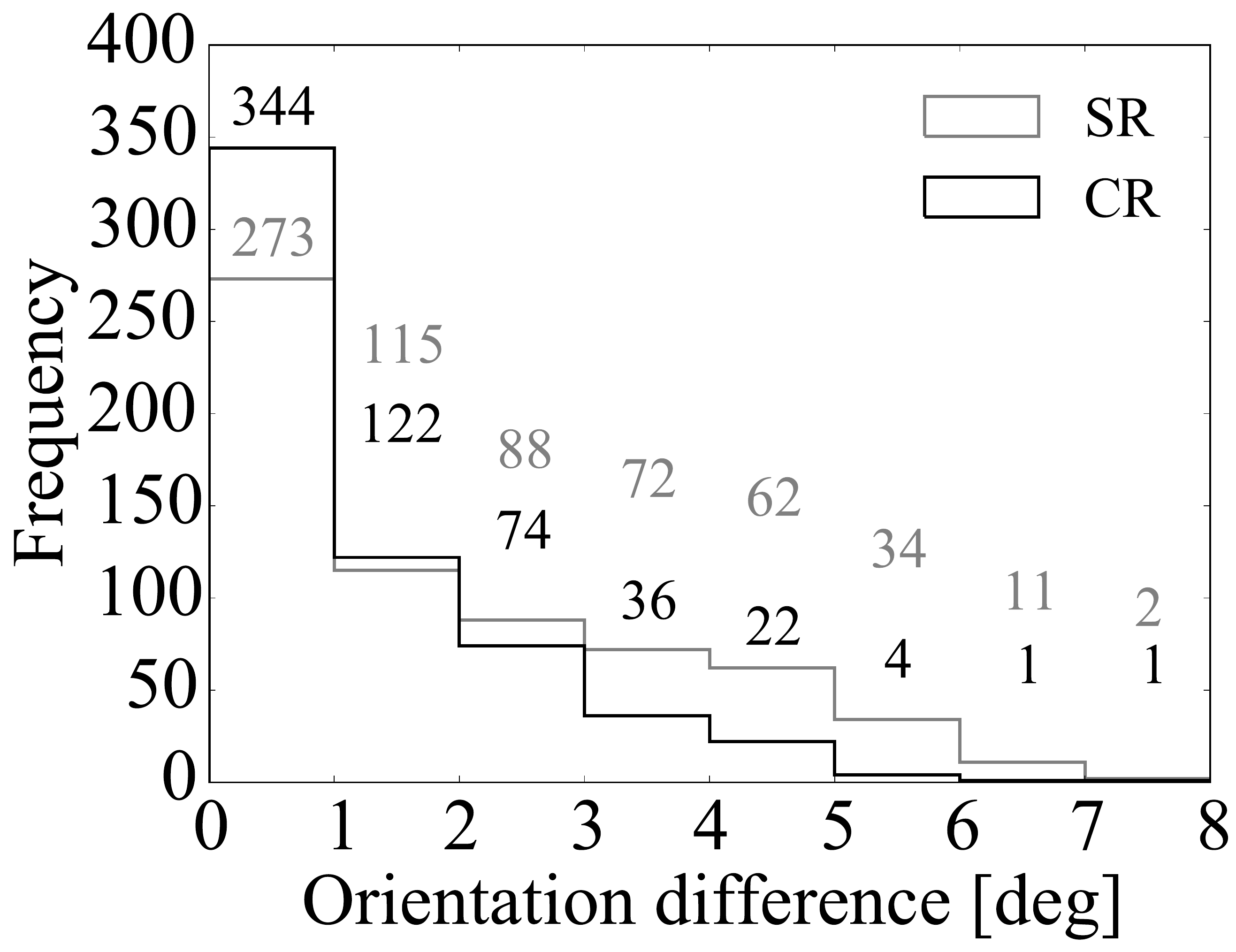}
   \end{figure}

A difference map was calculated for each region between the Sobel-gradient and manually measured orientation maps, and these are shown in~\autoref{fig:difference_map}. The histograms of the difference maps for both regions are shown in~\autoref{fig:difference_hist}. The majority of orientations differed by less than one degree. The maximum difference for the SR was 7.1$^{\circ}$, the mean difference was 1.9$^{\circ}$, and the standard deviation was 1.8$^{\circ}$, while that for the CR were 7.2$^{\circ}$, 1.2$^{\circ}$, and 1.3$^{\circ}$ respectively. \\

When measuring orientation manually, the non-linear skeletal path was essentially decomposed into small linear sections. The Sobel-gradient method does not decompose the path in this way, rather having a smooth transition along the filament that better reflects the curvature of the filaments path. The human defined section does not always align perfectly with the corresponding section in the Sobel map, sometimes they are shifted off each other by 1-2\,pixels. The larger orientation differences in the difference map mostly occur at locations where these shifts exist. This indicates that the larger difference values in the distribution in~\autoref{fig:difference_hist} are likely caused by this affect. This issue is unavoidable in constraining the uncertainty on the Sobel-gradient method in a realistic scenario--there is currently no other method on this scale besides manual measurement to provide a comparison.\\

The maximum of the difference maps of 7$^{\circ}$ is therefore a poor measure of the actual uncertainty of the Sobel-gradient method. It is more appropriate to use the standard deviation of the combined difference distribution of the SR and CR of $\sim$2$^{\circ}$ to estimate this uncertainty. This value is in agreement with that obtained from the circle test analysis. We therefore conservatively estimate the uncertainty of the Sobel-gradient method to be $\sim$2$^{\circ}$. This is acceptable considering that for Vela C BLASTPol data of \citet{fissel2016} the average uncertainty of the magnetic field maps is $\sim$2$^{\circ}$, reaching up to $\sim$16$^{\circ}$ in places.\\

The Sobel-gradient method is slightly more accurate than manual measurement, but it's strength of course is that it is significantly faster. To measure the orientation of one skeleton (such as those presented here) on this small scale manually takes most of a working day. When the additional time to then input the manual orientation measurement into a FITS file is taken into account, the manual process takes about one to two working days per skeleton. In studies of filaments, the filament and field relative orientation and radial profile measurements that involve filament orientation, are often repeated for hundreds of skeletons (all corresponding to the same input image, but to different combinations of input parameters to the filament identification algorithm). Consequently a fast and accurate orientation measurement method is essential. The Sobel-gradient method allows automation of the filament orientation measurement and is therefore crucial in this era of `big data'.  \\ 

\section{Applications of the Sobel-gradient method}
\label{sec:applications}
The Sobel-gradient method described is a technique to derive the orientation of filaments from their skeletons. This measurement has a number of astrophysical applications. One of the most significant is its use to calculate the relative orientation between magnetic fields and filaments, which provides clues on the role of magnetic fields in the formation and stability of filaments. We perform this calculation and present the results for the filaments of Vela C in Green et al. 2017 (in preparation). There are a number of other potential applications of the method including: to investigate relations between filament orientation and filament column density, mass, spatial width or molecular linewidth. 

\section{Summary} \label{sec:summary}
We have described a fully automated method to derive the orientation of a filament skeleton from the direction of the image intensity gradient that is suitable for complex, `looping' filamentary structures. We call this the `Sobel-gradient method'. It allows a local measurement of filament orientation that reflects the often rapid changes in orientation as a filament curves. This means that the filament orientation calculated from the intensity gradient can be directly compared to a map of the magnetic field, giving a quantitative, local measure of relative orientation as opposed to the qualitative, global and `by-eye' technique that is the current predominantly adopted method. It also has a number of other applications in investigating relationships involving filament orientation, such as that between filament orientation and column density. We have found this method to have a high degree of accuracy, with an uncertainty of $\sim$2$^{\circ}$. This computer vision technique provides the significant advantage that it can be easily automated, saving a significant amount of time compared to manual measurement, which is imperative in this era of `big data'. It also has broader applications and can be applied to any image containing lines or edges to find their orientation.\\





\acknowledgements
\noindent \textbf{Acknowledgements} \\
The authors would like to thank the referee Erik Rosolowsky and the anonymous statistical editor for their helpful comments which improved this work. The authors are also grateful to Eric Koch for helpful discussions surrounding the \textsc{filfinder} filament skeletons of this work. Herschel is an ESA space observatory with science instruments provided by European-led Principal Investigator consortia and with important participation from NASA. C.-E.G gratefully acknowledges the support of the Layne Beachley Foundation. L.M.F. and G.N. acknowledge support from NASA grant NNX13AE50G and from the Center for Interdisciplinary Exploration and Research in Astrophysics. L.M.F. was supported in part by an NSERC Postdoctoral Fellowship. LMF is a Jansky Fellow of the National Radio Astronomy Observatory (NRAO). NRAO is a facility of the National Science Foundation (NSF) operated under cooperative agreement by Associated Universities, Inc. This research made use of NASA's Astrophysics Data System. \\

\software{This project utilised  
\textsc{astropy} (\href{http://www.astropy.org}{http://www.astropy.org}) \citep{astropy2013}, \textsc{aplpy} (\href{http://aplpy.github.com}{http://aplpy.github.com}), \textsc{filfinder} \citep{koch2015}, Karma visualisation tools \citep{gooch1996}, and \textsc{scipy} (\href{http://www.scipy.org/}{http://www.scipy.org/}).
}


\end{document}